\documentclass[MSNbibl,nameyear,dvips]{arxstspdf}
\usepackage{tikz}
\usepackage{graphicx}
\usetikzlibrary{snakes,arrows,shapes}
\usepackage{url,breakurl}
\usepackage{flushend}
\usepackage{stfloats}

\volume{29}
\issue{4}
\pubyear{2014}
\firstpage{512}
\lastpage{528}
\doi{10.1214/14-STS490} 
\docsubty{FLA}

\makeatletter

\newtheorem{fact}{Fact}
\newtheorem{theorem}{Theorem}
\newtheorem{conjecture}{Conjecture}

\newcommand{\obs}{\mathrm{obs}}
\newcommand{\Bin}{\operatorname{Bin}}
\renewcommand{\epsilon}{\varepsilon}

\def\sfrac#1#2{#1/#2}

\makeatother

\begin{document}
\begin{frontmatter}

\title{Statistics, Causality and Bell's Theorem}
\runtitle{Statistics, Causality and Bell's Theorem}

\begin{aug}
\author[A]{\fnms{Richard D.} \snm{Gill}\corref{}\ead[label=e1]{gill@math.leidenuniv.nl}\ead[label=u1,url]{http://www.math.leidenuniv.nl/\textasciitilde gill}}
\runauthor{R. D. Gill}

\affiliation{University of Leiden}

\address[A]{Richard D. Gill is Professor, Mathematical Institute,
University of Leiden, Niels Bohrweg 1, Leiden, 2333 CA, The Netherlands
\printead{e1,u1}.}
\end{aug}

%
\begin{abstract}
Bell's [\textit{Physics} \textbf{1} (1964) 195--200] theorem is popularly supposed to establish the
nonlocality of quantum physics. Violation
of Bell's inequality in experiments such as that of Aspect, Dalibard and Roger [\textit{Phys. Rev. Lett.}
\textbf{49} (1982) 1804--1807] provides empirical proof of nonlocality in the real world.
This paper reviews recent work on Bell's theorem, linking it to issues in
causality as understood by statisticians. The paper starts with a
proof of a strong, finite sample, version of Bell's inequality and thereby
also of Bell's theorem, which states that quantum theory is
incompatible with the conjunction of three
formerly uncontroversial physical principles, here referred to as \emph
{locality},
\emph{realism} and \emph{freedom}.

Locality is the principle that the direction of causality matches the
direction of
time, and that causal influences need time to propagate spatially.
Realism and freedom are directly connected to statistical thinking on
causality: they relate to counterfactual reasoning, and to randomisation,
respectively. Experimental loopholes in state-of-the-art Bell type experiments
are related to statistical issues of post-selection in observational studies,
and the missing at random assumption.
They can be avoided by properly matching the statistical analysis to the
actual experimental design, instead of by making untestable assumptions of
independence between observed and unobserved variables.
Methodological and statistical issues in the design of quantum Randi
challenges (QRC)
are discussed.

The paper argues that Bell's theorem (and its experimental confirmation)
should lead us to relinquish not locality, but realism.
\end{abstract}

%
\end{frontmatter}

\section{Introduction}\label{sec1} Bell's (\citeyear{bell64})
theorem states that certain predictions of quantum mechanics are
incompatible with the conjunction of three fundamental principles
of classical physics which are sometimes given the short names
``realism'', ``locality'' and ``freedom''. Corresponding real world
experiments, Bell experiments, are supposed to demonstrate that this
incompatibility is a property not just of the theory of quantum
mechanics, but also of nature itself. The consequence is that we are
forced to reject at least one of these three principles.

Both theorem and experiment hinge around an inequality constraining
probability distributions of outcomes of measurements on
spatially separated physical systems; an inequality which must hold if
all three fundamental principles are true. In a nutshell, the inequality
is an empirically verifiable consequence of the idea that the outcome of
one measurement on one system cannot depend on which measurement
is performed on the other. This idea, called \emph{locality} or,
more precisely, \emph{relativistic local causality}, is just one of the
three principles. Its formulation refers to outcomes of measurements
which are not actually performed, so we have to assume their existence,
alongside of the outcomes of those actually performed: the principle of
\emph{realism}, or more precisely, \emph{counterfactual definiteness}.
Finally, we need to assume that we have complete \emph{freedom} to choose
\emph{which} of several measurements to perform---this is the third
principle, also called the \emph{no-conspiracy} principle or
\emph{no super-determinism}. (As we shall see, super-determinism
is a conspiratorial form of determinism.)
%
\begin{figure*}[b]

\includegraphics{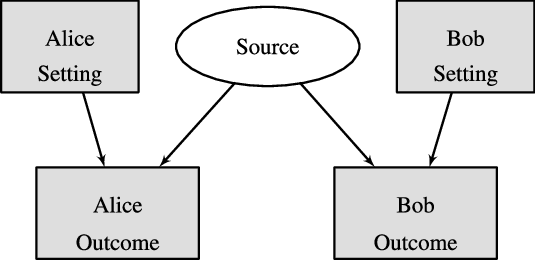}

\caption{A classical description of a Bell-CHSH type experiment entails the validity of the graphical model
described by this simple causal graph. Rectangles: observed variables; ellipse: unobserved. Settings and outcomes are both binary.
Experimental results arguably (via Bell's theorem) show that the classical description has to be abandoned.
The probability distribution of experimental data is far outside the class of probability distributions allowed by the model.}\label{fig1}
\end{figure*}

We shall implement the freedom assumption as the assumption of
statistical independence between the randomisation in a randomised
experimental design, and the set of outcomes of each experimental unit
under all possible treatments. This set consists of the
``counterfactual'' outcomes
of those treatments which were not actually applied,
as well as the ``factual'' outcome belonging to the treatment
chosen by the randomisation.

By existence of the outcomes of not actually performed
experiments, we mean their mathematical existence within some
mathematical-physical theory of the phenomenon in question.
So ``realism'' actually refers to models of reality, not to reality itself.
Moreover, it could be thought of as a somewhat idealistic position.
If we already have an adequate mathematical
physical model of reality, there would not seem to be a pressing need
to add into this theory some mathematical description of outcomes of experiments
which are not performed; and even if we do that, why should we demand
that these counterfactual objects satisfy the same kind of physical constraints
as the factual objects? However, it is a fact that prior to quantum physics,
realism was a completely natural property of all physical theories.

The concepts of realism
and locality together are often considered as one principle called
\emph{local realism}. Local realism is implied by the existence of
\emph{local hidden variables}, whether deterministic or stochastic. In a
precise mathematical sense, the reverse implication is also true: local
realism implies that we can construct a local hidden variable (LHV)
model for
the phenomenon under study. However, one likes to think of this assumption
(or pair of assumptions), the important thing to realize is that it is a
completely unproblematic feature of all classical physical theories;
freedom (no conspiracy) even more so.

The connection between Bell's theorem and statistical notions of
causality has
been noted many times in the past.
For instance, in a short note, \citet{robinsetal14}
derive Bell's inequality using the statistical language of \emph{causal
interactions}.
The \emph{causal graph} (DAG) of observed and unobserved variables
corresponding to a classical physical description of one run of a
standard Bell experiment
is given in Figure~\ref{fig1}.
Alice and Bob's settings are binary: they independently use
randomisation (a coin toss) to choose between one
of two settings on a measurement device. The outcome of each
measurement is also binary. Observed variables are
represented by grey rectangles; unobserved (there is only one,
but of course it might be of arbitrarily complex nature)
by a white oval.
The validity of this causal model places restrictions on the joint
distribution of the observed variables;
see, for instance, \citet{versteeggalstyan11}.

In view of the experimental support for violation of Bell's inequality,
the present writer prefers to imagine a world in which
``realism'' is not a fundamental principle of physics but only an
emergent property in the familiar realm of daily life.
In this way, we can keep quantum mechanics, locality and freedom.
This position does entail taking quantum randomness very seriously:
it becomes an irreducible feature of the physical world, a ``primitive notion'';
it is not ``merely'' an emergent feature. He believes that within this
position, the measurement problem (Schr\"odinger cat problem) has a
decent mathematical solution, in which causality is the guiding
principle (Slava Belavkin's ``eventum mechanics'').

Many practical minded physicists claim to be adherents
of the so-called Many Worlds interpretation (MWI)
of quantum mechanics. In the writer's opinion (but also of many writers on
quantum foundations), this interpretation also entails a rejection of
``realism'', but now in a very strong sense: the reality of
an actual random path taken by Nature through space--time is denied.
The only reality is the ensemble of all possible paths. Devilish experiments
lead to dead cats turning up on some paths, and alive cats on others.
According to MWI, the only reality is the quantum wave-function. The
reality of
the death (or not) of the cat is an illusion.

\section{Bell's Inequality}\label{sec2}

To begin with, I will establish a new version of the famous Bell
inequality (more precisely: Bell-CHSH inequality). My version is not an
inequality about theoretical expectation values, but is a probabilistic
inequality about
experimentally observed averages. Probability
derives purely from randomisation in the experimental design.

Consider a spreadsheet containing an $N\times4$ table of numbers $\pm
1$. The rows will be labelled by an index $j=1,\dots,N$. The columns are
labelled with names $A$, $A'$, $B$ and $B'$. I will denote the four
numbers in the $j$th row of the table by $A_j$, $A'_j$, $B_j$ and $B'_j$.
Denote by $\langle AB \rangle=(1/N)\sum_{j=1}^{N} A_j B_j$, the average
over the $N$ rows of the product of the elements in the $A$ and $B$
columns. Define $\langle AB' \rangle$, $\langle A'B \rangle$, $\langle
A'B' \rangle$ similarly.

Suppose that for each row of the spreadsheet, two fair coins are
tossed independently of one another, independently over all the rows.
Suppose that depending on the outcomes of the two coins, we either
get to see the value of $A$ or $A'$, and either the value of $B$ or $B'$.
We can therefore determine the value of just one of the four products
$AB$, $AB'$, $A'B$, and
$A'B'$, each with equal probability $\frac{1}4$,
for each row of the table. Denote by $\langle AB \rangle_{\obs}$ the
average of the observed products of $A$ and $B$ (``undefined'' if the sample
size is zero). Define $\langle AB' \rangle_{\obs}$, $\langle A'B
\rangle_{\obs}$ and $\langle A'B' \rangle_{\obs}$ similarly.

\begin{fact}\label{f2} For any four numbers $A$, $A'$, $B$, $B'$ each equal to
$\pm1$,
%
\begin{eqnarray}\label{eq1}
AB+AB'+A'B-A'B' = \pm2.
\end{eqnarray}
\end{fact}

\begin{pf}
Notice that
\[
AB+AB'+A'B-A'B' = A
\bigl(B+B'\bigr)+A'\bigl(B-B'\bigr).
\]
$B$ and $B'$ are
either equal to one another or unequal. In the former case, $B-B'=0$ and
$B+B'=\pm2$; in the latter case $B-B'=\pm2$ and $B+B'=0$. Thus,
$AB+AB'+A'B-A'B'$ equals either $A$ or $A'$, both of which equal
$\pm1$, times $\pm2$. All possibilities lead to $AB+AB'+A'B-A'B'=\pm
2$.
\end{pf}

\begin{fact}\label{f1}
%
\begin{eqnarray}\label{eq2}
\langle AB \rangle+\bigl\langle AB' \bigr\rangle+\bigl\langle
A'B \bigr\rangle-\bigl\langle A'B' \bigr
\rangle\le2.
\end{eqnarray}
\end{fact}

\begin{pf}
By (\ref{eq1}),
\begin{eqnarray*}
&&\langle AB \rangle+\bigl\langle AB' \bigr\rangle+\bigl\langle
A'B \bigr\rangle-\bigl\langle A'B' \bigr
\rangle
\\
&&\quad = \bigl\langle AB+AB'+A'B-A'B'
\bigr\rangle\in[-2,2].
\end{eqnarray*}
\upqed
\end{pf}

Formula (\ref{eq2}) is known as the CHSH inequality (\cite{clauseretal69}). It is a generalisation of the original
\citet{bell64} inequality.

When $N$ is large one would expect $\langle AB \rangle_{\obs}$ to be
close to $\langle AB \rangle$, and the same for the other
three averages of observed products. Hence, equation (\ref{eq2}) should remain
approximately true when we replace the averages of the four products over
all $N$ rows with the averages of the four products in each of four
disjoint sub-samples of expected size $N/4$ each.
The following theorem expresses this intuition in a precise and useful way.
Its straightforward proof, given in the \hyperref[app]{Appendix}, uses two \citet{hoeffding63}
inequalities (exponential bounds on the tail of binomial and
hypergeometric distributions) to
probabilistically bound the difference between
$\langle AB \rangle_{\obs}$ and $\langle AB \rangle$, etc.

\begin{theorem}\label{TH1} Given an $N\times4$ spreadsheet of numbers $\pm1$ with
columns $A$, $A'$, $B$ and $B'$, suppose that, completely at random, just
one of $A$ and $A'$ is observed and just one of $B$ and $B'$ are observed
in every row. Then, for any $\eta\ge0$,
%
\begin{eqnarray}\label{eq3}
&&\Pr \bigl( \langle AB \rangle_{\obs}+\bigl\langle AB' \bigr
\rangle_{\obs}+\bigl\langle A'B \bigr\rangle_{\obs}\nonumber\\
&&\hspace*{61pt} {}-
\bigl\langle A'B' \bigr\rangle_{\obs}\le 2 +
\eta \bigr)\\
&&\quad  \ge 1- 8 e^{- N ( \sfrac{\eta}
{16})^2 }.\nonumber
\end{eqnarray}
\end{theorem}

Traditional presentations of Bell's theorem derive the large $N$ limit
of this result.
If for $N\to\infty$, experimental averages converge to theoretical mean
values, then by (\ref{eq3}) these must satisfy
%
\begin{eqnarray}\label{eq4}
\langle AB \rangle_{\mathrm{lim}}+\bigl\langle AB' \bigr
\rangle_{\mathrm
{lim}}+\bigl\langle A'B \bigr\rangle_{\mathrm{lim}}-
\bigl\langle A'B' \bigr\rangle_{\mathrm{lim}} \le 2.\hspace*{-20pt}
\end{eqnarray}
Like (\ref{eq2}), this inequality is also called the CHSH inequality.

I conclude this section with an open problem.
An analysis by \citet{vongehr13} of
the original Bell inequality, which is ``just'' the CHSH inequality
in the situation that one of the four correlations is identically equal
to $\pm1$,
suggests that the following conjecture might be true. I come back to this
in the last section of the paper.

\begin{conjecture}\label{co1} Under the assumptions of Theorem~\ref{TH1},
%
\begin{eqnarray}\label{eq5}
&&\Pr \bigl( \langle AB \rangle_{\obs}+\bigl\langle AB' \bigr
\rangle_{\obs}+\bigl\langle A'B \bigr\rangle_{\obs}-
\bigl\langle A'B' \bigr\rangle_{\obs} > 2 \bigr)\nonumber\\[-7pt]\\[-9pt]
&&\quad \le \tfrac{1}2.\hspace*{-5pt}\nonumber
\end{eqnarray}
\end{conjecture}

\section{Bell's Theorem}\label{sec3}

Both the original Bell inequality, and
Bell-CHSH inequality (\ref{eq4}), can be used to prove \emph{Bell's theorem}:
quantum mechanics is incompatible with the principles of realism,
locality and freedom. If we want to hold on to all three
principles, quantum mechanics must be rejected. Alternatively, if we
want to hold on to quantum theory, we have to relinquish at least one of
those three principles.

An \emph{executive summary} of the proof of Bell's theorem
consists purely of the following one-liner:
certain models in quantum physics, referring to an experiment
with the layout of Figure~\ref{fig1}, predict
%
\begin{eqnarray}\label{eq6}
&&\langle AB \rangle_{\mathrm{lim}}+\bigl\langle AB' \bigr
\rangle_{\mathrm
{lim}}+\bigl\langle A'B \bigr\rangle_{\mathrm{lim}}-
\bigl\langle A'B' \bigr\rangle_{\mathrm{lim}}\nonumber\\[-8pt]\\[-8pt]
&&\quad  = 2\sqrt
2.\nonumber
\end{eqnarray}
More details will be given in a moment.

If we accept quantum mechanics, should we reject locality, realism, or freedom?
Almost no-one is prepared to abandon freedom. It seems to be a matter of
changing fashion whether one blames locality or realism.
I will argue that we must place the blame on realism, and not in the weak
sense of the Copenhagen interpretation which is a kind of dogmatic
assertion that it doesn't make any sense to ask
``what is actually going on behind the scenes'', but in a more positive
sense: the positive assertion that quantum randomness is both
real and fundamental.
In classical physics, randomness is \emph{merely} the result of
dependence on uncontrollable initial conditions.
Variation in those conditions, or uncertainty about them, leads
to variation, or uncertainty, in the final result.
However, there is no such explanation for quantum randomness.
Quantum randomness is intrinsic, nonclassical, irreducible.
It is not an emergent phenomenon. It is the bottom line.
It is a fundamental feature of the fabric of reality.

For present purposes, we do not need to understand any of the quantum mechanics
behind (\ref{eq6}): we just need to know the specific statistical predictions
which follow
from a particular model in
quantum physics called the EPR-B model. The initials refer here to the
celebrated paradox of \citet{einsteinetal35}
in a version introduced by \citet{bohm51}. The
EPR-B model
predicts the statistics of measurements of spin on each of an entangled
pair of
spin-half quantum systems in the singlet state. Fortunately, we do not
need to understand
any of these words in order to understand what an EPR-B experiment
looks like
(see Figure~\ref{fig1} again).

In one run of this stylised experiment, two particles are generated
together at a source, and then travel to two distant locations.
Here, they are measured by two experimenters Alice and Bob. Alice and Bob
are each in possession of a measurement apparatus which can ``measure the
spin of a particle in any chosen direction''. Alice (and similarly, Bob)
can freely choose (and set) a \emph{setting} on her measurement
apparatus. Alice's setting is an arbitrary direction in real
three-dimensional space represented by a unit vector $\mathbf a$. Her
apparatus will then register an observed outcome $\pm1$ which is called
the observed spin of Alice's particle in direction $\mathbf a$. At the
same time, far away, Bob chooses a direction $\mathbf b$ and also gets to
observe an outcome $\pm1$. This is repeated many times---the complete
experiment will consist of a total of $N$ runs. We will imagine Alice and
Bob repeatedly choosing new settings for each new run, in the same
fashion as in Section~\ref{sec2}: each tossing a fair coin to make a binary choice
between just two possible settings, $\mathbf a$ and $\mathbf a'$ for
Alice, $\mathbf b$
and $\mathbf b'$ for Bob.

First, we will complete our description of the
quantum mechanical predictions for each run separately.
For pairs of particles in the singlet state, the prediction of quantum
mechanics is that in whatever directions Alice and Bob perform their
measurements, their outcome $\pm1$ is completely
random, that is, both marginal distributions are uniform. The outcomes are
not, however, independent. They are correlated, with correlation
depending on the
two settings. To be precise, the expected value of the product of the outcomes
is equal to $-\mathbf a \cdot\mathbf b=-\cos(\theta)$
where $\theta$ is the angle between the two directions.

With this information, we can write down the $2 \times2$ table
for the joint probability distribution of the outcomes
at the two locations, given two settings differing in direction by the
angle $\theta$:\vspace*{12pt}

\centerline{
\begin{tabular}{@{}lcc@{}}
\hline
& $\boldsymbol{+}\mathbf{1}$ & $\boldsymbol{-}\mathbf{1}$ \\ \hline
$+1$ & ${ \frac{1}4} (1-\cos(\theta))$ & ${ \frac{1}4}
(1+\cos(\theta))$ \\[3pt]
$-1$ & ${ \frac{1}4} (1+\cos(\theta))$ & ${ \frac{1}4} (1-\cos(\theta))$ \\
\hline
\end{tabular}
}\vspace*{12pt}

\noindent Both marginals of the table are uniform. The expectation of the product
of the outcomes equals the probability that they are equal
minus the probability they are different
$\frac{2}4(1-\cos(\theta))-\frac{2}4 (1+\cos(\theta))=-\cos(\theta)$. Physicists
use the word ``correlation'' to refer to the raw (uncentered, unnormalised)
product moment but in this case the physicist's and the statistician's
correlation coincide.

As mentioned before, Alice and Bob now perform $N$ runs of the
experiment according to the following randomised experimental design.
Alice has fixed in
advance two particular directions $\mathbf a$ and $\mathbf a'$; Bob has
fixed in
advance two particular directions $\mathbf b$ and $\mathbf b'$.
In each run, Alice and Bob are each sent one of a
new pair of particles in the singlet state. While their particles are
\emph{en route} to them, they each toss a fair coin in order to choose
one of their two measurement directions. In total $N$ times, Alice
observes either $A=\pm1$ or $A'=\pm1$ say, and Bob observes either
$B=\pm1$ or $B'=\pm1$. At the end of the experiment, four
``correlations'' are calculated: the four sample averages
of the products $AB$, $AB'$, $A'B$ and $A'B'$. Each correlation is based
on a different subset of runs, of expected size $N/4$, determined by the
$2N$ fair coin tosses.

Under \emph{realism} we can imagine, for each run, alongside of the
outcomes of the actually measured pair of variables, also the outcomes of
the not measured pair. Under \emph{locality}, the outcomes in Alice's wing
cannot depend on the choice of which variable is measured in Bob's wing.
Thus, for each run there is a suite of potential outcomes $A$, $A'$,
$B$ and
$B'$, but only one of $A$ and $A'$, and only one of $B$ and $B'$ actually
gets to be observed. By \emph{freedom}, the choices are statistically
independent of the actual values of the four.

I will assume furthermore that the suite of counterfactual outcomes in the
$j$th run does not actually depend on which particular variables were
observed in previous runs.
This \emph{memoryless} assumption can be completely avoided by
using the martingale version of Hoeffding's inequality, \citet{gill03}.
But the present analysis is already applicable if we imagine $N$
copies of the experiment each with only a
single run, all being done simultaneously in different laboratories.

The assumptions of realism, locality and freedom have put us firmly
in the situation of the previous section. Therefore, by Theorem~\ref{TH1}, the
four sample correlations (empirical raw product moments) satisfy (\ref{eq3}).
%
\begin{figure}[b]
\centerline{\begin{tikzpicture}
     \draw[->] (0,0) -- (canvas polar cs:angle=0,radius=1cm) node[right] {$\mathbf a$};
     \draw[->] (0,0) -- (canvas polar cs:angle=90,radius=1cm)
                                                 node[above] {$\mathbf a^\prime$};
     \draw[->] (0,0) -- (canvas polar cs:angle=135,radius=1cm)
                                                 node[above] {$\mathbf b^\prime$};
     \draw[->] (0,0) -- (canvas polar cs:angle=225,radius=1cm) node[below] {$\mathbf b$};
\end{tikzpicture}}
\caption{Four measurement directions, all in the same plane. Alice's
settings are the two orthogonal directions $\mathbf a$, $\mathbf a'$,
and Bob's settings are the orthogonal $\mathbf b$, $\mathbf b'$.
Relative to one another, the pairs are arranged so that $\mathbf a'$
and $\mathbf b'$ are close to pointing in the same direction, while at
the same time the other three pairs of one of Alice's and one of Bob's
settings ($\mathbf a$ and $\mathbf b$, $\mathbf a$ and $\mathbf b'$,
$\mathbf a'$ and $\mathbf b$) are all equally close to pointing in
opposite directions.}\label{fig2}
\end{figure}
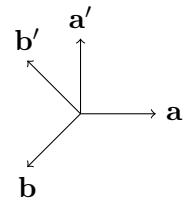

Let us contrast this prediction with the quantum mechanical predictions
obtained with a certain clever selection of directions. We will take the
four vectors $\mathbf a$, $\mathbf a'$, $\mathbf b$ and $\mathbf b'$ to
lie in the
same plane. It is then enough to specify the angles $\alpha$,
$\alpha'$, $\beta$, $\beta'\in[0,2 \pi]$ which they make with respect
to some
fixed vector in this plane. Consider the choice $\alpha=0$,
$\alpha'=\pi/2$, $\beta=5\pi/4$, $\beta'=3\pi/4$; see Figure~\ref{fig2}.
The differences
$|\alpha-\beta|$, $|\alpha-\beta'|$, $|\alpha'-\beta|$ are all equal to
$\pi\pm\pi/4$: these pairs of angles are only $45$ degrees away from
being opposite to one
another; the corresponding measurements are quite strongly positively
correlated. On the other hand, $|\alpha'-\beta'|=\pi/4$: these two
angles are $45$ degrees away from being equal and the corresponding
measurements are as
strongly anti-correlated, as the other pairs were strongly
correlated.
Three of the correlations are equal to $-\cos(3\pi/4)=
-(-1/\sqrt2)=1/\sqrt2$ and the fourth is equal to
$-\cos(\pi/4)=-1/\sqrt2$. Thus, we would expect to see, up to statistical
variation,
\begin{eqnarray*}
&&\langle AB \rangle_{\obs}+\bigl\langle AB' \bigr
\rangle_{\obs}+\bigl\langle A'B \bigr\rangle_{\obs}-
\bigl\langle A'B' \bigr\rangle_{\obs}\\
&&\quad \approx4/
\sqrt2 = 2 \sqrt2 \approx 2.828,
\end{eqnarray*}
cf. (\ref{eq6}). By Tsirelson's inequality (\cite{tsirelson80}), this is actually
the largest absolute deviation from the CHSH inequality which is allowed
by quantum mechanics.

Many experiments have been performed confirming these predictions.
Two particularly notable ones are those of \citet{aspectetal82} in
Orsay, Paris, and of \citet{weihsetal98} in Innsbruck (later I will discuss two of the most recent).

In these experiments, the choices of which
direction to measure were not literally made with coin tosses,
but by rather more practical physical systems.
In Alain Aspect's Orsay and Gregor Weihs' Innsbruck experiments,
the separation between the locations of Alice and Bob was large; the
time it took from
initiating the choice of random direction to measure to completion of the
measurement was small: so small, that Alice's measurement is
complete before a signal traveling at the speed of light could possibly
transmit Bob's choice to Alice's location. However, note that this
depends on what one
considers to be the time each randomisation starts happening. Weihs'
experiment improves on Aspect's
in this respect.

The data gathered from the Innsbruck experiment is available online. It
had $N\approx15\mbox{,}000$; and found $\langle AB \rangle_{\obs}+\langle
AB' \rangle_{\obs}+\langle A'B \rangle_{\obs}-\langle A'B'
\rangle_{\obs} = 2.73\pm0.022$, the statistical accuracy (standard
deviation) following from a standard delta-method calculation assuming
i.i.d. observations per setting pair. The reader can check that this
corresponds to accuracy obtained by a standard computation using
binomial variances of the counts for each of the four roughly equal sub-samples.
By (\ref{eq3}), under realism, locality and freedom, the chance that $\langle AB
\rangle_{\obs}+\langle AB' \rangle_{\obs}+\langle A'B
\rangle_{\obs}-\langle A'B' \rangle_{\obs}$ would exceed $2.73$
is less than $10^{-12}$.

The experiment deviates in several ways from what has been described so
far, and I will summarise them here.

An unimportant difference is the physical system used: polarisation of
entangled photons rather than spin of entangled spin-half particles
(e.g., electrons).

An important difference between the
idealisation and the truth concerns the picture of Alice and Bob
repeating some actions $N$ times with $N$ fixed in advance.
The experimenters do not control when a pair of
photons will leave the source nor how many times this happens.
Even talking about ``pairs of photons'' is using classical physical language
which can be acutely misleading. In actual fact, all we observe are
individual detection events (time, current setting, outcome)
at each of the two detectors, that is, at each
measurement apparatus.

Complicating this still further is the fact that
many particles fail to be detected at all. One
could say that the outcome of measuring one particle is not binary but
ternary: $+$, $-$, or \emph{no detection}. If neither particle of
a pair is detected, then we do not even know there is a pair at all.
$N$ was not only not fixed in advance: it is not even known.
The data cannot be summarised in a list of pairs of settings
and pairs of outcomes (whether binary of ternary), but consists of two
lists of the \emph{random times} of definite measurement outcomes in
each wing
of the experiment together with the settings in force at the time of the
measurements. The settings are being extremely rapidly, randomly switched,
between the two alternative values.
When detection events occur close together in time they are treated as
belonging to a pair of photons.

In Weihs' experiment, only $1$ in $20$ of the events in each wing of
the experiment seemed to be paired with an event in the other. If all
detections correspond
to emissions of pairs from the source, then for every $400$ pairs of
photons, just one pair leads to a paired event, $2\times19$ lead to unpaired
events, and the remaining $361$ to no observed event at all.

We will return to the issue of whether the idealised picture of $N$ pairs
of particles, each separately being measured, each particle in just one
of two ways, is really appropriate, in a later section; we will also
take a look then at two more, very recent, experiments. However, the
point is that quantum mechanics does seem to promise that experiments of
this nature could in principle be done, and if so, there seems no reason
to doubt they could violate the CHSH inequality. Three correlations more
or less equal to $1/\sqrt2$ and one equal to $-1/\sqrt2$ have been
measured in the lab. Not to mention that the whole curve $-\cos(\theta)$
has been experimentally recovered.

Right now the situation is that
at least four major experimental groups
(Singapore, Brisbane, Vienna, Illinois)
seem to be vying to be the first to perform a successful
and completely ``loophole-free'' experiment,
predictions being that this is no more than five years away
(cf. Marek \.Zukowski, quoted in \cite{merali10}).
It will be a major achievement, the crown of more than fifty years' labour.

\section{Realism, Locality, Freedom}\label{sec4}

This section and the next are about metaphysics and can safely be skipped
by the reader impatient to learn more about statistical aspects of Bell
experiments.

The EPR-B correlations have a second message beyond the fact that they
violate the CHSH inequality. They also exhibit perfect anti-correlation
in the case that the two directions of measurement are exactly equal---and
perfect correlation in the case that they are exactly opposite. This
brings us straight to the EPR argument not for the nonlocality of
quantum mechanics, but for the incompleteness of quantum mechanics.

\citet{einsteinetal35}
were revulsed by the idea that the
``last word'' in physics would be a ``merely'' statistical theory.
Physics should explain why, in each individual instance, what actually
happens does happen. The belief that every ``effect'' \emph{must} have a
``cause'' has driven Western science since Aristotle. Now according to
the singlet correlations, if Alice were to measure the spin of her
particle in direction $\mathbf a$, it is certain that if Bob were to do the
same, he would find exactly the opposite outcome. Since it is
inconceivable that Alice's choice has any immediate influence on the
particle over at Bob's place, it must be that the outcome of measuring
Bob's particle in the direction $\mathbf a$ is predetermined ``in the
particle'' as it were. The measurement outcomes from measuring spin in
all conceivable directions on both particles must be predetermined
properties of those particles. The observed correlation is merely caused
by their origin at a common source.

Thus Einstein used locality, together with the predictions of quantum
mechanics itself, to infer realism or
counterfactual definiteness in the strong sense that the outcomes of
measurements on physical systems are predefined properties of those
systems, carried in them, and merely revealed by the act of measurement.
From this, he argued the \emph{incompleteness}
of quantum mechanics---it describes some aggregate
properties of collectives of physical systems, but does not even deign to
talk about physically definitely existing properties of individual
systems.

Whether it needed external support or not, the notion of counterfactual
definiteness is nothing strange in all of physics (prior to the
invention of quantum mechanics). It comes for free with a deterministic
view of
the world as a collection of objects blindly obeying definite rules.

Instead of assuming quantum mechanics and deriving counterfactual definiteness,
Bell turned the EPR argument on its head. He assumes three principles which
Einstein would have endorsed anyway, and uses them to get a
contradiction with
quantum mechanics; and the first is counterfactual definiteness.
We must first agree that though, say, only $A$ and $B$ are actually
measured in
one particular run, still, in a mathematical sense, $A'$ and $B'$ also exist
(or at least may be constructed) alongside of the other two;
and moreover, they may be thought to be located in space and time
just where one would imagine. Only after that does
it make sense to discuss \emph{locality}:
the assumption that \emph{which} variable is
being observed at Alice's location does not influence the values taken by
the other two at Bob's location.

Having assumed realism and locality, we can bring the freedom
assumption into play.
As we have seen, it allowed us to analyse our experiment with classical
probability
tools based on a classically randomised design.
Some writers like to associate the freedom assumption with the free
will of the experimenter, others with the existence of ``true''
randomness in other physical processes: either way, one metaphysical assumption
is justified by another. I~would rather see it in a practical way: we understand
pseudo-randomness very well, and its principles underly coin tossing
just as much as pseudo random generators. We use randomisation effectively
in all kinds of contexts (randomised algorithms, randomised clinical
trials, randomised designs).
Do we really want to believe that the observed correlations $\pm
0.7071$ (three positive,
one negative) come about through a physical mechanism
by which the outcomes of two coin tosses and two polarisation measurements
are all exquisitely dependent on one another through \emph{all four}
being jointly \emph{pre-determined} by events in the deep past?
When such a hypothesis is otherwise completely unnecessary? (Otherwise,
we never see spatial-temporal correlations following this sign pattern,
larger in absolute value than $0.5$).
A mechanism which is completely unknown? A~mechanism which ensures
in effect that Alice's photon knows how Bob's photon is being measured?
Yet, a mechanism which cannot make any of those correlations larger in
absolute value than $0.7071$,
though if it really were the case that Alice's photon knows Bob's
setting, three positive
and one negative correlation $\pm1.0$ could have been achieved.

I think that Occam's razor tells us to discard this flavour of
super-determinism, also known
as conspiracy. In fact, to abandon freedom means to abandon science:
we may discard all empirical (observational) data. Everything is explained
but nothing can be predicted.

Keeping freedom, we have to make a choice between two other inconceivable
possibilities: do we reject locality, or do we reject realism?

Here, I would like to call on Occam's principle again. Suppose realism is
true. Instead of invoking the fact that a collection of four
coin toss outcomes and photo-detector
clicks were jointly predetermined in the deep past, we now have to invoke
instantaneous communication across large distances of the outcomes
of these processes, by as yet unknown processes, and again
with only the extremely subtle and special effects which quantum
mechanics seems to predict. Alice cannot communicate with Bob
through this phenomenon. There is no observable action-at-a-distance.
The surface predictions of quantum mechanics are perfectly compatible
with relativistic causality. It is only when we hypothesise a hidden layer
that we run into difficulties.

It seems to me that we are pretty much forced into rejecting
\emph{realism}, which, please remember, is actually an idealistic
concept: outcomes ``exist'' of measurements which were \emph{not}
performed. However, I admit it goes against all instinct. In the case of
equal settings, how can it be that the outcomes are equal and opposite,
if they were not predetermined at the source?

Though it is perhaps only a comfort blanket,
I would like here to appeal to the limitations of our own
brains, the limitations we experience in our ``understanding'' of physics
due to our own rather special position in the universe. In philosophy,
this notion is called \emph{embodied cognition}. There is also hard
empirical evidence for this idea.

According to cognitive scientists
(see, for instance, \cite{spelkekinzler07}),
our brains are at birth hardwired
with various basic conceptions about the world. These ``modules'' are
called \emph{systems of core knowledge}. The idea is that we
cannot acquire new knowledge from our sensory experiences (including
learning from experiments: we cry, and food and/or comfort is provided)
without having a prior framework in which to interpret the data of
experience and experiment. It seems that we have modules for
algebra and modules for geometry: basic notions of number and of space. Most
interestingly in the present context, we
also have modules for causality. We distinguish between \emph{objects}
and \emph{agents} (we learn that we ourselves are agents). Objects are
acted on by agents. Objects have continuous existence in space--time, they
are local. Agents can act on objects, also at a distance. Together this
seems to me to be a built-in assumption of determinism: we have been
created (by evolution) to operate in an Aristotelian world, a world in
which every effect has a cause.

The argument (from physics, and by Occam's razor, not from
neuroscience) for abandoning realism is made eloquently by Boris Tsirelson
in an internet encyclopaedia article on entanglement
(\href{http://en.citizendium.org/wiki/Entanglement_\%28physics\%29}
{Citizendium: entanglement}).
It was Tsirelson from whom I borrowed the terms counterfactual definiteness,
relativistic local causality, and no-conspiracy. He points out that it is
a mathematical fact that quantum physics is consistent with relativistic
local causality and with no-conspiracy. In all of physics, there is no
evidence \emph{against} either of these two principles.

I would like to close this section by mentioning a beautiful paper
by \citet{masanesetal06}
who argue in a very general setting
(i.e., not assuming quantum theory, or local realism, or anything) that
\emph{quantum nonlocality}, by which they mean the violation of Bell
inequalities, together with \emph{nonsignalling}, which is the property
that the marginal probability distribution seen by Alice of $A$ does not
depend on whether Bob measures $B$ of $B'$, together imply
indeterminism: that is to say: that the world is stochastic, not
deterministic.

\section{Resolution of the Measurement Problem}\label{sec5}

The measurement problem, also known as Schr\"odin\-ger's cat problem, is
the problem of how to reconcile two apparently mutually contradictory
parts of
quantum mechanics. When a quantum system is isolated from the rest of the
world, its quantum state (a vector, normalised to have unit length, in
Hilbert space) evolves unitarily, deterministically. When we look at a
quantum system from outside, by making a measurement on it in a
laboratory, the state collapses to one of the eigenvectors of an operator
corresponding to the particular measurement, and it does so with
probabilities equal to the squared lengths of the projections of the
original state vector into the eigenspaces. Yet the system being measured
together with the measurement apparatus used to probe it form together a
much larger quantum system, supposedly evolving unitarily and
deterministically in time.

Accepting that quantum theory is intrinsically sto\-chastic, and accepting
the reality of measurement outcomes, led \citet{belavkin00} to a
mathematical framework which he called eventum mechanics which (in my
opinion) indeed reconciles the two faces of quantum physics
(Schr\"odinger evolution, von Neumann collapse) by a most simple device.
Moreover, it is based on ideas of causality with respect to time. I have
attempted to explain this model in as simple terms as possible in
\citet{gill09}. The following words will only make sense to those with some
familiarity with quantum mechanics.

The idea is to model the world in the conventional way
with a Hilbert space, a quantum state on that space, and a
unitary evolution. Inside this framework, we look for a collection of
bounded operators on the Hilbert space which all commute with one
another, and which are \emph{causally} compatible with the unitary
evolution of the space, in the sense that they all commute with past
copies of themselves (in the Heisenberg picture, one thinks of the
quantum observables as changing, the state as fixed; each observable
corresponds to a time indexed family of bounded operators). We call this
special family of operators the \emph{beables}: they correspond to
physical properties in a classical-like world which can co-exist, all
having definite values at the same time, and definite values in the past
too. The state and the unitary evolution together determine a joint
probability distribution of these time-indexed variables, that is, a
stochastic process. At any fixed time, we can condition the state of the
system on the past trajectories of the beables. This leads to a quantum
state over all bounded operators which commute with all the beables.

The result is a theory in which the deterministic and stochastic parts of
traditional quantum theory are combined into one harmonious
whole. In fact, the notion of restricting attention to a sub-class of all
observables goes back a long way in quantum theory under the name
\emph{super-selection rule}; and abstract quantum theory (and quantum
field theory) has long worked with arbitrary algebras of observables, not
necessarily the full algebra of a specific Hilbert space. With respect to
those traditional approaches the only novelty is to suppose that the
unitary evolution when restricted to the sub-algebra is not invertible.
It is an endomorphism, not an isomorphism. There is an arrow of time.

It turns out that the theory is mathematically equivalent to important versions
of the continuous spontaneous localisation (CSL) model, a way to solve
the measurement problem by adding an explicit stochastic collapse term to
the Schr\"odinger equation (Initially, the two theories seem quite
different in
nature). The problem of crafting a relativistically invariant version
of CSL remained
open for many years (and was a major obstruction to its acceptance) yet just
recently this problem has been solved by \citet{beddingham11}.
See Pearle (\citeyear{pearle97}, \citeyear{pearle12}) for
further details.

CSL has been eloquently championed over the years by Philip Pearle and
I refer the
reader to his many works, in particular the two just cited, both
explaining CSL
and explaining why it does solve the measurement problem, while MWI
does not.

\section{Loopholes}\label{sec6}

In real world experiments, the ideal experimental protocol of particles
leaving a source at definite times, and being measured at distant
locations according to locally randomly chosen settings cannot be
implemented.

Experiments have been done with pairs of entangled ions, separated only
by a short distance. The measurement of each ion takes a relatively long
time, but at least it is almost always successful. Such experiments are
obviously blemished by the so-called \emph{communication} or
\emph{locality} loophole. Each particle can know very well how the other
one is being measured.

Many very impressive experiments have been performed with pairs of
entangled photons. Here, the measurement of each photon can be performed
very rapidly and at huge distance from one another. However, many photons
fail to be detected at all. For many events in one wing of the
experiment, there is often no event at all in the other wing, even though
the physicists are pretty sure that almost all detection events do
correspond to (members of) entangled pairs of photons. This is called the
\emph{detection} loophole. Popularly it is thought to be merely connected
to the efficiency of photo-detectors and that it will be easily overcome
by the development of better and better photo-detectors. Certainly that is
necessary, but not sufficient, as I will explain.

In Weihs' experiment mentioned earlier, only 1 in 20 of the events in
each wing of the experiment is paired with an event in the other wing.
Thus, of every 400 pairs of photons---if we assume
that detection and nondetection occur independently of one
another in the two wings of the experiment---only 1 pair results in a
successful measurement of both the photons; there are 19 further unpaired
events in each wing of the experiment; and there were 361 pairs of
photons not observed at all.

Imagine (anthropocentrically) classical particles\linebreak[4]  about to leave the
source and aiming to fake the singlet correlations. If they are allowed
to go undetected often enough, they can engineer any correlations they
like, as follows. Consider two new photons about to leave the source.
They agree between one another with what pair of settings they would like
to be measured. Having decided on the desired setting pair, they next
generate outcomes $\pm1$ by drawing them from the joint probability
distribution of outcomes given settings, which they want the experimenter
to see. Only then do they each travel to their corresponding detector.
There, each particle compares the setting it had chosen in advance with
the setting chosen by Alice or Bob. If they are not the same, it decides
to go undetected.
With probability 1$/$4 we will have successful detections in both wings of
the experiment. For those detections, the pair of settings according to
which the particles are being measured is identical to the pair of
settings they had aimed at in advance.

This example illustrates that if one wants to experimentally \emph{prove}
a violation of local realism without making
the untestable statistical assumption of ``missing at random'',
known as the \emph{fair-sampling assumption} in this context,
one has to put
limits on the amount of ``nondetection''.
There is a long history and big literature on this topic.
I will just mention one of such results.

Larsson (\citeyear{larsson98}, \citeyear{larsson99})
has proved variants of the CHSH
inequality which take account of the possibility of nondetections. The
idea is that under local realism, as the proportion of ``missing''
measurements increases from zero, the upper bound ``2'' in the CHSH
inequality (\ref{eq4}) increases, too. We introduce a quantity $\gamma$ called the
\emph{efficiency of the experiment}: this is the minimum over all setting
pairs of the probability that Alice sees an outcome given Bob sees an
outcome (and vice versa). It is not to be confused with ``detector
efficiency''. It turns out that the (sharp) bound on $\langle AB \rangle
_{\mathrm{lim}}+\langle
AB' \rangle_{\mathrm{lim}}+\langle A'B \rangle_{\mathrm{lim}}-\langle A'B'
\rangle_{\mathrm{lim}}$ set by local realism is no longer $2$ as in
(\ref{eq4}), but
$2+\delta$, where $\delta=\delta(\gamma)= 4(\gamma^{-1} -1)$.

As long as $\gamma\ge1/\sqrt2 \approx0.7071$,
the bound $2+\delta$ is smaller than $2 \sqrt2$. Weihs' experiment has
an efficiency of 5\%. If only we could increase it to above 71\%
\emph{and} simultaneously keep the state and measurements close to
perfection, we could have definitive experimental proof of Bell's
theorem.

This would be correct for a ``clocked'' experiment. Suppose now particles
determine themselves the times that they are measured. Thus, a local
realist pair of particles trying to fake the singlet correlations could
arrange between themselves that their measurement times are delayed by
smaller or greater amounts depending on whether the setting they see at
the detector is the setting they want to see, or not.
It turns out that this gives our devious particles even more scope for
faking correlations. \citet{larssonandgill04} called
this the \emph{coincidence loophole}, and derived the sharp bound on
$\langle AB \rangle_{\mathrm{lim}}+\langle AB' \rangle_{\mathrm
{lim}}+\langle A'B
\rangle_{\mathrm{lim}}-\langle A'B' \rangle_{\mathrm{lim}}$ set by
local realism
is $2+\delta$, where now $\delta=\delta(\gamma)= 6(\gamma^{-1} -1)$. As
long as $\gamma\ge3(1-1/\sqrt2) \approx0.8787$,
the bound $2+\delta$ is smaller than $2 \sqrt2$. We need to get
experimental efficiency above 88\%, and keep everything else close to
perfect at
the very limits allowed by quantum physics.

How far is there still to go? In 2013, the Vienna group published a
paper in
the journal \textit{Nature} entitled
``Bell violation using entangled photons without the fair-sampling assumption''
(\cite{giustinaetal13}).
The authors write ``this is the very first time that an experiment has been
done using photons which does not suffer from the detection loophole'', and
moreover, the experiment ``makes the photon the first physical system
for which each of the main loopholes has been closed, albeit in
different experiments''.

It was however rapidly pointed out that the experiment was actually vulnerable
to the coincidence loophole, not ``just'' to the detection loophole.
Now, it actually
should be possible to simply re-analyse the data from that experiment,
defining coincidences
with respect to an externally defined lattice of time intervals instead
of relative
to observed detection times only. Ideally, this will only slightly
increase the
``singles rate'' and slightly decrease the number of coincidences,
thereby slightly decreasing both size and statistical significance of
the Bell violation,
but hopefully without altering the substantive conclusion. A more stringent
re-analysis of the data (\cite{larssonetal13})
has confirmed that the initial claims were justified.

In the meantime, exploiting this gap between results and claims, a consortium
led by researchers from Illinois have published their own experimental results,
also reporting that theirs is ``the first experiment that fully closes
the detection loophole with photons, which are then the only system
in which both loopholes have been closed, albeit not simultaneously''
(\cite{christensenetal13}).
They used the \citet{larssonandgill04} inequality.

Is this just a question of prestige? No: various new quantum
technologies depend on quantum entanglement,
and in particular, various cryptographic communication protocols are
not secure as long as it is
possible to ``fake'' violation of Bell inequalities with classical systems.

It is \emph{logically possible} that quantum mechanics itself could
prevent one ever
from performing a both successful and loophole-free experiment.
Quantum uncertainty relations could in principle prevent the creation
of a
multipartite quantum system, whose components can be measured in well-separated
space--time regions, while simultaneously those components
are in the required joint entangled state.
I christened this possibility ``Bell's fifth position'' in \citet{gill03}.
Here, I just mention that the possibility had already been championed
for many years by
Emilios Santos, whose paper \citet{santos05} is well
worth reading.

On the other hand, continuous improvement of experimental techniques over
more than fifty years has seen continuous pushing of detection efficiency
toward the critical boundaries, without any attenuation of the quantum
correlations of the singlet state. Now that we are getting very close indeed
to the boundary, it would seem very unlikely that we won't be able to
go past it.

I conclude this section with mention of some recent work on the
conspiracy loophole.
Recently, \citet{galliccioetal14} have made the
novel suggestion to rule out conspiracy by the experimental device of
choosing settings
with the help of detection times of photons arriving from widely separated,
and very distant galaxies from the dawn of time.
This idea will probably be implemented soon in an experiment by
Zeilinger. I am not however convinced by
this idea: though Alice's setting choice is triggered by a photon which
cannot yet have interacted
with Bob's measurement device or the source, still the setting itself
is also partially determined
by Alice's detection apparatus, and it certainly has. And if there is
dependence between subsequent
settings on Alice's side, then on Bob's side it soon becomes possible
to predict future settings
(this is known as the memory loophole).

In my opinion, we have to rule out conspiracy (ensure freedom) by
choosing settings by a
cascade of classical randomness (coins, pseudo RNGs, etc.). We can
never logically
rule out conspiratorial super determinism, but we can make appeal to
this escape clause ludicrous.

\section{Bell's Theorem Without Inequalities}\label{sec7}

In recent years new proofs of Bell's theorem have been invented which
appear to avoid probability or statistics altogether, such as the famous
GHZ (Greenberger, Horne, Zeilinger) proof. Experiments have already been
done implementing the set-up of these proofs, and physicists have claimed
that these experiments prove quantum---nonlocality by the outcomes of a
finite number of runs: no statistics, no inequalities (yet their papers
do exhibit error bars).

Such a proof runs along the following lines. Suppose local realism is
true. Suppose also that some event $\mathcal A$ is certain. Suppose that
it then follows from local realism that another event $\mathcal B$ has
probability zero, while under quantum mechanics it can be arranged that
the same event $\mathcal B$ has probability one. Paradoxical, but not a
contradiction in terms: the catch is that events $\mathcal A$ and
$\mathcal B$ are events under different experimental conditions: it is
only under local realism and freedom that the events $\mathcal A$ and
$\mathcal B$ can be situated in the same sample space. Freedom
is needed here to equate the probabilities of observable events with
those of
unobservable events, just as in our own proof of Bell's theorem.
We need to be able to assume that the subset of
runs of the experiment in which the events were observable
are a random sample of the set of all repetitions.

When we use randomisation in experimental design, we are assuming that
randomisation is independent of pre-existing unobserved characteristics
of the experimental units. Is it plausible that the outcomes of coin
tosses used to
create a randomised experimental design were predetermined together with
properties of the experimental units under study, so that our random sub-samples
of units being given a particular treatment, are actually heavily
biased with regard
to the properties we measure on them? Of course, under a purely deterministic
world view (super-determinism) everything that was ever going to happen
was determined in advance, at the dawn of creation. But even in a
deterministic world,
pseudo-randomness exists and is well understood.

As an example, consider the following scenario, generalizing the
Bell-CHSH scenario to the situation where the outcome of the measurements
on the two particles is not binary, but an arbitrary real number. This
situation has been studied by \citet{zohrenandgill08},
\citet{zohrenetal10}.

Just as before, settings are chosen at random in the two wings of the
experiment. Under local realism we can introduce variables $A$, $A'$, $B$
and $B'$ representing the outcomes (real numbers) in one run of the
experiment, both of the actually observed variables, and of those not
observed.

It turns out that it is possible under quantum mechanics to arrange that
$\Pr\{B'\le A\}=\Pr\{A\le B\}=\Pr\{B\le A'\}=1$ while $\Pr\{B'\le
A'\}=0$. On the other hand, under local realism, $\Pr\{B'\le
A\}=\Pr\{A\le B\}=\Pr\{B\le A'\}=1$ implies $\Pr\{B'\le A'\}=1$.

Note that the four probability measures under which, under quantum
mechanics, $\Pr\{A \le B \}$, $\Pr\{A \ge B' \}$, $\Pr\{A' \ge B \}$,
$\Pr\{A' \ge B' \}$ are defined, refer to four different experimental
set-ups, according to which of the four pairs $(A,B)$, etc. we are
measuring.

The experiment to verify these quantum mechanical predictions has not yet
been performed though some colleagues are interested. Interestingly,
though it requires a quantum entangled state, that state should
\emph{not} be the maximally entangled state (the amount of
entanglement of a state can be quantified in many ways, for instance
through entropy notions, but it would take us too far into the quantum
formalism to explain that here). Maximal ``quantum
nonlocality'' is quite different from maximal entanglement. And this is
not an isolated example of the phenomenon.

Note that even if the experiment is repeated a large number of times, it
can never prove that probabilities like $\Pr\{A\le B\}$ are exactly equal
to 1. It can only give strong statistical evidence, at best, that the
probability in question is very close to 1 indeed. But actually
experiments are never perfect and more likely is that after a number of
repetitions, one discovers that $\{A>B\}$ actually has positive
probability---that event will happen a few times. The experimenter cannot
create the required quantum state exactly, measurements are not perfect.
Thus, the logical conclusion from the experiment is that nothing has
been proved.

To be sure, one can give a proof of Bell's theorem that the theory of
quantum mechanics is in conflict with local realism, which relies only
on logic,
not on probability. But if we want to use the set-up of the proof as a
set-up for
an experiment, we move to a different ball-park. We want to perform an
experiment
which gives strong evidence that nature is incompatible with local realism.
It turns out that whatever experimental set-up we take, we will necessarily
find ourselves explicitly or implicitly in the business of
statistically proving
violation of inequalities, as the next section will make clear.

\section{Better Bell Inequalities}\label{sec8}

Why all the attention to the CHSH inequality? There are others around,
aren't there? And are there alternatives to ``inequalities'' altogether?
I will argue here that the whole story is ``just'' a collection
of inequalities, and the reason behind this can be expressed
in a simple geometric picture.

In a precise sense, the CHSH inequality is the only Bell inequality worth
mentioning in the scenario of two parties, two measurements per party,
two outcomes per measurement. Let us generalise this scenario and consider
$p$ parties, each choosing between one of $q$ measurements, where each
measurement has $r$ possible outcomes (further generalisations are
possible to unbalanced experiments, multi-stage experiments, and so on).
I want to explain why CHSH plays a very central role in the $2 \times2
\times2$ case, and why in general, \emph{generalised Bell inequalities}
are all there is when studying the $p\times q\times r$ case. The short
answer is: these inequalities are the bounding hyperplanes of a convex
polytope of ``everything allowed by local realism''. The vertices of the
polytope are deterministic local realistic models. An arbitrary local
realist model is a mixture of the models corresponding to the vertices.
Such a mixture is a \emph{hidden variables model}, the hidden variable
being the particular random vertex chosen by the mixing distribution in a
specific instance.

From quantum mechanics, after we have fixed a joint $p$-partite quantum
state, and sets of $q$ $r$-valued measurements per party, we will be able
to write down probability tables $p(a,b,\ldots|x,y,\ldots)$ where the variables
$x$, $y$, etc. take values in $1,\dots,q$, and label the measurement used
by the first, second, \dots party. The variables $a$, $b$, etc., take
values in $1,\dots,r$ and label the possible outcomes of the
measurements. Altogether, there are $q^p r^p$ ``elementary
probabilities'' in this list of tables. More generally, any
specific instance of a theory, whether local-realist, quantum mechanical,
or beyond, generates such a list of probability tables, and defines
thereby a point in $q^p r^p$-dimensional Euclidean space.

We can therefore envisage the sets of all local-realist models, all
quantum models, and so on, as subsets of $q^p r^p$-dimensional Euclidean
space. Now, whatever the theory, for any values of $x$, $y$, etc., the
sum of the probabilities $p(a,b,\dots|x,y,\dots)$ must equal $1$. These
are called \emph{normalisation constraints}. Moreover, whatever the
theory, all probabilities must be nonnegative: \emph{positivity
constraints}. Quantum mechanics is certainly local
in the sense that the marginal distribution of
the outcome of any one of the measurements of any one of the parties does
not depend on which measurements are performed by the other
parties. Since marginalization corresponds again to summation of
probabilities, these so-called \emph{no-signalling constraints} are
expressed by linear equalities in the elements in the probability tables
corresponding to a specific model. Not surprisingly, local-realist models
also satisfy the no-signalling constraints.
%
\begin{figure*}

\includegraphics{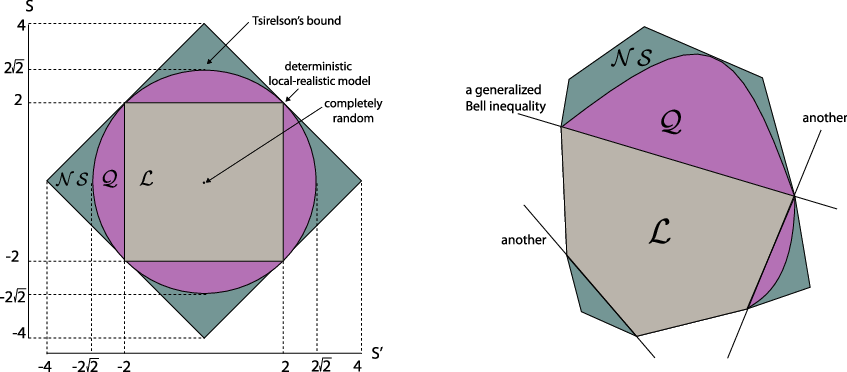}

\caption{Left: a caricature of the $2\times2\times2$ case. It
actually lives in 8, not 2 dimensions. Right: caricature of the general
case in which (bottom left) a further possibility is allowed: no purple
between the green and grey. Artwork: Daniel Cavalcanti.}\label{fig3}
\end{figure*}

We will call a list of probability tables restricted only by positivity,
normalisation and no-signalling, but otherwise completely arbitrary, a
\emph{no-signalling} model. The positivity constraints are linear inequalities
which place us in the positive orthant of Euclidean space. Normalisation
and no-signalling are linear equalities which place us in a certain
affine sub-space of Euclidean space. Intersection of orthant and affine
sub-space creates a convex polytope: the set of all no-signalling models.
We want
to study the sets of local-realist models, of quantum models, and of
no-signalling models. We already know that local-realist and quantum are
contained in no-signalling. It turns out that these sets are
successively larger,
and strictly so: \emph{quantum} includes all \emph{local-realist} and
more (that's Bell's theorem); \emph{no-signalling} includes all \emph{quantum}
and more (that is Tsirelson's inequality combined with an example of a
no-signalling model which violates Tsirelson's inequality).

Let us investigate the local-realist models in more detail. A special
class of local-realist models are the \emph{local-deterministic} models.
A local-deterministic model is a model in which all of the probabilities
$p(a,b,\dots|x,y,\dots)$ equal $0$ or $1$ \emph{and} the no-signalling
constraints are all satisfied. This implies that for each
possible measurement by each party, the outcome is
prescribed, independently of what measurements are made by the other
parties. Now, it is easy to see that any local-realist model corresponds
to a probability mixture of local-deterministic models. After all, it
``is'' a joint probability distribution of simultaneous outcomes of each
possible measurement on each system, and thus it ``is'' a probability
mixture of degenerate distributions: fix the random element $\omega$, and
each outcome of each possible measurement of each party is fixed; we
recover their joint distribution by picking $\omega$ at random.

This makes the set of local-realist models a convex polytope: all
mixtures of a finite set of extreme points. Therefore, it can also be
described as the intersection of a finite collection of half-spaces, each
half-space corresponding to a boundary hyperplane.

It can also be shown that the set of quantum models is closed and convex,
but its boundary is very difficult to describe.

Let us think of these three models from ``within'' the affine sub-space of
no-signalling and normalisation. Relative to this sub-space, the no-signalling
models form a full (nonempty interior) closed convex polytope. The
quantum models form a strictly smaller closed, convex, full set. The
local-realist models form a strictly smaller still, closed, convex, full
polytope.

Slowly, we have arrived at a rather simple picture,
Figure~\ref{fig3}.
Imagine a square, with a circle inscribed in it,
and with another smaller square inscribed within the circle.
The outer square represents the boundary of the set of
all no-signalling models. The circle is the boundary of the convex set
of all
quantum models. The square inscribed within the circle is the boundary of
the set of all local-realist models. The picture is oversimplified. For
instance, the vertices of the local-realist polytope are also extreme
points of the quantum body and vertices of the no-signalling polytope.

A generalised Bell inequality is simply a boundary hyperplane, or face,
of the local-realist polytope, relative to the normalisation and
no-signalling affine sub-space, and excluding boundaries corresponding to
the positivity constraints. I will call these interesting boundary
hyperplanes ``nontrivial''. In the $2\times2\times2$ case, for which
the affine sub-space where all the action lies is 8-dimensional, the
local-realist polytope has exactly 8 nontrivial boundary hyperplanes.
They correspond exactly to all possible CHSH inequalities (obtained by
permuting outcomes, measurements and parties). Thus, in the $2\times
2\times2$ case, the Bell-CHSH inequality is indeed ``all there is''.

When we increase $p$, $q$ or $r$, new Bell inequalities turn up, and
moreover, keep turning up (``new'' means not obtainable from ``old'' by
omitting parties or measurements or grouping outcomes). It seems a
hopeless (and probably pointless) exercise to try to classify them.

A natural question is whether every nontrivial generalised Bell inequality
can actually be violated by quantum mechanics. I posed this as an
open question a long time ago, and for a long time the answer seemed
probably be ``yes''. However, a nice counter-example has recently
been discovered; see \citet{almeidaetal10}.

Quite a few generalised Bell inequalities have turned out to be of
particular interest, for instance, the work of Zohren and Gill concerned
the $2\times2\times r$ case and discussed a class of inequalities, one
for each $r$, whose asymptotic properties could be studied as $r$
increased to infinity. Further statistical connections to
missing data problems and optimal experimental design,
have been exploited by \citet{vandametal05}
and \citet{gill07}.

Much of the material of this section is covered in an excellent
survey paper by \citet{brunneretal13},
from which I took, with the permission both of the authors and of
the artist Daniel Cavalcanti, the two illustrations in Figure~\ref{fig3}:
the first is a cartoon of the $2\times2\times2$ case, the second of
the general case.

\section{Quantum Randi Challenges}\label{sec9}

A second reason for the specific form of the proof of Bell's theorem
which started this paper is that it lends itself well to design of
computer challenges. Every year, new researchers publish, or try to
publish, papers in which they claim that Bell made some fundamental
errors, and in which they put forward a specific local realist model
which allegedly reproduces the quantum correlations. The papers are long
and complicated; the author finds it hard to get the work published, and
suspects a conspiracy by ``The Establishment''. The claims regularly
succeed in attracting media attention, occasionally becoming head-line
news in
serious science journalism; some papers are published, too, and
not only in obscure journals.

Extraordinary claims require extraordinary evidence.
I used to find it useful in debates with ``Bell-deniers'' to challenge them
to implement their local realist model as computer programs for a network
of classical computers, connected so as to mimic the time and space
separations of the Bell-CHSH experiments.

The protocol of the challenge I issued in the past is the following.
Bell-denier is to write computer programs for three personal
computers, which are to play the roles of source $\mathcal S$,
measurement station $\mathcal A$, and measurement station $\mathcal B$.
The following is to be repeated say 15\mbox{,}000 times. First, $\mathcal S$
sends messages to $\mathcal A$ and $\mathcal B$. Next, connections
between $\mathcal A$, $\mathcal B$ and $\mathcal S$ are severed. Next,
from the outside world so to speak, I deliver the results of two coin
tosses (performed by myself), separately of course, as input setting to
$\mathcal A$ and to $\mathcal B$. Heads or tails correspond to a request
for $A$ or $A'$ at $\mathcal A$, and for $B$ or $B'$ at $\mathcal B$. The
two measurement stations $\mathcal A$ and $\mathcal B$ now each output an
outcome $\pm1$. Settings and outcomes are collected for later data
analysis, Bell-denier's computers are re-connected; next run.

Bell-denier's computers can contain huge tables of random numbers, shared
between the three, and of course they can use pseudo-random number
generators of any kind. By sharing the pseudo-random keys in advance,
they have resources to any amount of shared randomness they like.

In \citet{gill03}, I showed how a martingale Hoeffding
inequality
gives an exponential bound like (\ref{eq3}) in the situation just
described. This enabled me to choose $N$, and a criterion for win/lose
(say, halfway between $2$ and $2 \sqrt2$), and a guarantee to
Bell-denier (at least so many runs with each combination of settings),
such that I would happily bet 3000 Euros any day that Bell-denier's
computers will fail the challenge.

The point (for me) was not to win money for myself, but to enable the
Bell-denier who considers accepting the challenge (a personal challenge
between the two of us, with adjudicators to enforce the protocol) to
discover for him or herself that ``it cannot be done''. It is important
that the adjudicators do not need to look inside the programs written by
the Bell-denier, and preferably do not even need to look inside his
computers. They are black boxes. The only thing that has to be enforced
are the communication rules. However, there are difficulties here. What
if Bell-denier's computers are using a wireless network which the
adjudicators cannot detect?

A new kind of computer challenge, called the ``quantum Randi challenge'',
was proposed in 2011 by Sascha Vongehr
(\href
{http://www.science20.com/alpha_meme/official_quantum_randi_challenge-80168}{Science2.0:
QRC}).
It is inspired by the well known challenge to ``paranormal phenomena''
by James Randi
(scientific sceptic and fighter against pseudo-science, see
\shref{http://en.wikipedia.org/wiki/James\_Randi}{Wikipedia: James\\Randi}).
Vongehr's challenge
(see Vongehr, \citeyear{vongehr12}, \citeyear{vongehr13})
differs in a number of fundamental respects from
mine, which indeed was not a quantum Randi
challenge in Vongehr's sense.

Sascha Vongehr's QRC completely
cuts out any necessity for communication, protocol verification, adjudication.
In fact, the Bell-denier no longer has to cooperate with myself or with any
other member of the establishment. They simply have to write a program
which should perform a certain task. They post their program on internet.
If others find that it does indeed perform that task, the news will
spread like wildfire.

Vongehr prefers Bell's original inequality, and I prefer CHSH, so I will
here present an (unauthorised) ``CHSH style'' modification of his QRC.

Suppose someone has invented a local hidden variables theory.
He can use it to simulate $N=800$ runs of a CHSH experiment. Typically,
he will simulate the source, the photons, the detectors, all in one program.
Let us suppose
that his computer code produces reproducible results, which means that
the code or the application is reasonably portable, and
will give \emph{identical} output when run on another computer with the
same inputs.
In particular, if it makes use of a pseudo random number generator (RNG),
it must have the usual ``save'' and ``restore'' facilities for the seed
of the RNG.
Let us suppose that the program calls the RNG the same number
of times for each run, and that the program does not make use in any way
of memory of past measurement settings. The program must accept any
legal stream of
pairs of binary measurement settings of any length~$N$.

In particular then, the program can be run with $N=1$ and all four possible
pairs of measurement settings, and the same initial random seed,
and it will thereby generate successively four pairs $(A,B)$,
$(A',B)$, $(A,B')$, $(A',B')$. If the programmer neither cheated nor
made any errors,
in other words, if the program is a correct implementation of a
genuine LHV model, then
both values of $A$ are the same, and so are both values of $A'$, both
values of $B$, and both values of $B'$.
We now have the first row of the $N\times4$ spreadsheet of Section~\ref{sec2}
of this paper.

The random seed at the \emph{end} of the previous phase is now used as
the initial seed
for another phase, the second run, generating a second row of the
spreadsheet. This is where the
prohibition of exploiting memory comes into force. The second row of
counterfactual
outcomes has to be completed without knowing which particular setting
pair Alice and
Bob will actually pick for the first row.

Notice that the LHV model is allowed to use \emph{time}, since the
saved random seeds
could also include the current run number and the initial random seed
value, too: in other words,
when doing the calculations for the $n$th run, the LHV model has access
to everything
it did in the previous $n-1$ runs.

My claim is that a correct implementation of a bona-fide LHV model
which does not exploit the memory loophole can be used to fill in
the $N\times4$ spreadsheet of Section~\ref{sec2}. When we now generate
random settings and calculate the correlations, we get the same results
as if they had been submitted in a single stream to the same program,
run once with the same
initial seed.

My new CHSH-style QRC to any local realist out there who is interested,
is that they program their LHV model, modified so that
it simply accepts a random seed and value of $N$, and outputs an
$N\times4$ spreadsheet.
They should post it on internet and draw attention to it on any of the
many internet fora devoted to
discussions of quantum foundations.
Anyone interested runs the program, generates $N\times2$ settings, and
calculates CHSH.
If the program reproducibly, repeatedly (significantly more than half
the time, cf. Conjecture \ref{co1} of Section~\ref{sec2}),
violates CHSH,
then the creator has created a classical physical system which
systematically violates the CHSH inequalities,
thereby disproving Bell's theorem.
No establishment conspiracy can stop this news from spreading round the world,
everyone can replicate the experiment.
The creator will get the Nobel prize and there will be incredible
repercussions throughout physics.

Some local realists will however insist on using memory.
They cannot rewrite their programs to create one $N\times4$ spreadsheet.
Instead, $N$ rounds of communication are needed between themselves and
some trusted neutral vetting agency.
To borrow an idea I learnt from Han Geurdes,
we should think of some kind of rating agency such as those for banks,
an independent agency which carries out ``stress tests'', on demand,
but at a reasonable price, to anyone who is interested and will pay.
The procedure is almost as before: it ensures yet again that the LHV
model is legitimate,
or more precisely, is legitimate in its implemented form.
The agency generates a first run of settings (i.e., one setting pair),
but keeps it secret for the moment.
The LHV theorist supplies a first run-set of values of $(A,A',B,B')$.
The agency reveals the first setting pair, the LHV theorist generates a
second run set $(A,A',B,B')$.
This is repeated $N=800$ times.
The whole procedure can be repeated any number of times,
the results are published on internet,
everyone can judge for themselves.

\begin{appendix}
\section*{Appendix: Proof of Theorem~\texorpdfstring{\protect\ref{TH1}}{1}}\label{app}
The proof of (\ref{eq3}) will use the following two Hoeffding inequalities:

\begin{fact}[(Binomial)]\label{f3}
Suppose $X\sim\Bin(n,p)$ and $t>0$.
Then
\[
\Pr(X/n \ge p+t) \le \exp\bigl(- 2n t^2\bigr).
\]
\end{fact}

\begin{fact}[(Hypergeometric)]\label{f4}
Suppose $X$ is the number of red balls found in a sample without
replacement of size $n$ from a vase containing $pM$ red balls and
$(1-p)M$ blue balls and $t>0$.
Then
\[
\Pr(X/n \ge p+t) \le \exp\bigl(- 2n t^2\bigr).
\]
\end{fact}

\begin{pf*}{Proof of Theorem~\protect\ref{TH1}}
In each row of our $N\times4$ table of numbers $\pm1$, the product
$AB$ equals $\pm1$.
For each row, with probability $1/4$, the product is either observed or
not observed.
Let $N_{AB}^{\obs}$ denote the number of rows in which both $A$ and
$B$ are observed.
Then $N_{AB}^{\obs}\sim\Bin(N,1/4)$, and hence by Fact~\ref{f3}, for
any \mbox{$\delta>0$},
\[
\Pr \biggl(\frac{N_{AB}^{\obs}}{N} \le {\frac{1} 4} - \delta \biggr) \le \exp
\bigl(- 2N \delta^2\bigr).
\]

Let $N_{AB}^+$ denote the total number of rows (i.e., out of $N$) for
which $AB=+1$, define $N_{AB}^-$ similarly.
Let $N_{AB}^{{\obs},+}$ denote the number of rows such that $AB=+1$
among those selected for observation of $A$ and $B$.
Conditional on $N_{AB}^{\obs}= n$, $N_{AB}^{{\obs},+}$ is
distributed as the number of red balls in a sample without replacement of
size $n$ from a vase containing $N$ balls of which $N_{AB}^+$ are red
and $N_{AB}^-$ are blue. Therefore by Fact~\ref{f4}, conditional on
$N_{AB}^{\obs}= n$, for any $\epsilon>0$,
\[
\Pr \biggl(\frac{N_{AB}^{{\obs},+}}{N_{AB}^{{\obs}}}\ge \frac{N_{AB}^{+}}{N}+\epsilon \biggr)\le\exp\bigl(-
2n \epsilon^2\bigr).
\]

Recall that $\langle AB \rangle$ stands for the average of the
product $AB$ over the whole table; this can be rewritten as
\[
\langle AB \rangle= \frac{N_{AB}^{+}-N_{AB}^{-}}{N} = 2 \frac{N_{AB}^{+}}{N} -1.
\]
Similarly, $\langle AB \rangle_{\obs}$ denotes the average of the
product $AB$ just over the rows of the table for which both $A$ and $B$
are observed; this can be rewritten as
\[
\langle AB \rangle_{\obs} = \frac{N_{AB}^{{\obs},+}-N_{AB}^{{\obs},-}}{
N_{AB}^{{\obs}}} = 2 \frac{N_{AB}^{{\obs},+}}{N_{AB}^{{\obs}}} -1.
\]
For given $\delta>0$ and $\epsilon>0$, all of $N_{AB}^{{\obs}}$,
$N_{AB'}^{{\obs}}$, $N_{A'B}^{{\obs}}$ and
$N_{A'B'}^{{\obs}}$ are at least $(\frac{1}4-\delta)N$ with probability at
least $1-4\exp(-2N\delta^2)$.
On the event where this happens, the conditional probability that
$\langle AB \rangle_{\obs}$ exceeds $\langle AB \rangle+2\epsilon$ is
bounded by
\[
\exp\bigl(- 2 N^{{\obs}}_{AB} \epsilon^2\bigr)\le
\exp\bigl(- 2 \bigl({\tfrac{1}{4}} - \delta\bigr)N \epsilon^2
\bigr).
\]
The same is true for the other three averages (for the last one we first
exchange the roles of $+$ and $-$ to get a bound on $\langle
-A'B'\rangle_{\obs}$).
Combining everything, we get that
\[
\langle AB \rangle_{\obs}+\bigl\langle AB' \bigr
\rangle_{\obs}+\bigl\langle A'B \bigr\rangle_{\obs}-
\bigl\langle A'B' \bigr\rangle_{\obs} \le 2 +8
\epsilon,
\]
except possibly on an event of probability at most
\[
p = 4\exp\bigl(-2N\delta^2\bigr)+4 \exp\bigl(- 2 \bigl({
\tfrac{1}{4}}-\delta\bigr)N \epsilon^2\bigr).
\]
We want to bound $p$ by $8\exp(-N(\eta/16)^2)$ where $\eta=8\epsilon$,
making $(\eta/16)^2=(\epsilon/2)^2$.
Choosing $8\delta^2=\epsilon^2$, we find $2\delta^2=(\epsilon/2)^2=(\eta/16)^2$.
If $ 8(\frac{1}4-\delta) \ge1$, then $p \le8 \exp(- 2N\delta^2)$ and we
are home.
The restriction on $\delta$ translates to $\delta\le\frac{1}8$ and thence
to $\eta\le2 \sqrt2$.
But for $\eta>2$, (\ref{eq3}) is trivially true anyway, so the restriction on
$\eta$ can be forgotten.
\end{pf*}

\end{appendix}

\section*{Acknowledgments}
I'm grateful to the anonymous referees and to Gregor Weihs, Anton
Zeilinger, Stefano Pironio,
Jean-Daniel Bancal, Nicolas Gisin, Samson Abramsky, and Sascha Vongehr
for ideas, criticism, references\ldots.
I especially thank Bryan Sanctuary, Han Geurdes and Joy Christian for
their tenacious and
spirited arguments \emph{against} Bell's theorem which motivated
several of the results
presented here.



\end{document}